%% file: main.tex
\documentclass[letterpaper]{article} 
\usepackage{aaai25}  
\usepackage{times}  
\usepackage{helvet}  
\usepackage{courier}  
\usepackage[hyphens]{url}  
\usepackage{graphicx} 
\usepackage{natbib}  
\usepackage{caption} 
\usepackage{algorithm}
\usepackage{algorithmic}

\usepackage{newfloat}
\usepackage{listings}
\DeclareCaptionStyle{ruled}{labelfont=normalfont,labelsep=colon,strut=off} 
\floatstyle{ruled}
\newfloat{listing}{tb}{lst}{}
\floatname{listing}{Listing}

\usepackage{xcolor}

\title{Reddit Rehab: User Migration in Response to Mobile Client Shutdowns}
\author {
    Franz Waltenberger\textsuperscript{\rm 1,2},
    Angelina Voggenreiter\textsuperscript{\rm 1},
    Martin Paul Wessel\textsuperscript{\rm 1,2},
    Juergen Pfeffer\textsuperscript{\rm 1}
}
\affiliations {
    \textsuperscript{\rm 1}Technical University of Munich, Munich, Germany\\
    \textsuperscript{\rm 2}Center for Digital Technology and Management, Munich, Germany\\
    waltenberger@cdtm.de
}

\begin{document}

\maketitle

\begin{abstract}
\input{sections/0_abstract}

\end{abstract}

\input{sections/1_introduction}
\input{sections/2_background}
\input{sections/3_data}

\input{sections/4_methods}

\input{sections/5_results}
\input{sections/6_discussion}

\bibliography{references}

\end{document}

%% file: sections/0_abstract.tex
This paper investigates the behavior of Reddit users who relied on alternative mobile apps, such as Apollo and RiF, before and after their forced shutdown by Reddit on July 1, 2023.
The announcement of the shutdown led many observers to predict significant negative consequences, such as mass migration away from the platform. Using data from January to November 2023, we analyze user engagement and migration rates for users of these alternative clients before and after the forced discontinuation of their apps. We find that 22\% of alternative client users permanently left Reddit as a result, and 45\% of the users who openly threatened to leave if the changes were enacted followed through with their threats. Overall, we find that the shutdown of third-party apps had no discernible impact on overall platform activity. While the preceding protests were severe, ultimately for most users the cost of switching to the official client was likely far less than the effort required to switch to an entirely different platform. 
Scientific attention has increased to understand the contributing factors and effects of migration between online platforms, but real-world examples with available data remain rare. Our study addresses this by examining a large-scale online migratory movement.

%% file: sections/1_introduction.tex
\section{Introduction}



Users of online platforms are often seen as "fleeting creatures", quickly spooked and, if not fully satisfied, already on their way to the next new thing. Past examples of companies losing users over sometimes seemingly insignificant design changes or management decisions appear to support this notion \cite{garcia2013social, newell2016user, franzmann2020mobile}. One such example was the redesign of the popular messaging app Snapchat, which lost three million daily active users following an unpopular redesign in early 2018 \cite{digitaltrends2018}. In this paper, we examine the effects of the forced shutdown of six alternative Reddit mobile clients on July 1st, 2023, adding to the existing discussions about the dynamics of user migration in online networks.
In April 2023, the management of the popular social media platform Reddit announced its plans to monetize access to the site's Application Programming Interface (API) \cite{announcement2024Apr}. According to Reddit's CEO, the main motivation for this decision was the rise of large language models, which until then had been trained for free on Reddit comment data \cite{Shakir2023Apr}. While on other platforms, most regular users would not even necessarily notice new API terms, this proved to be an unpopular decision among many Reddit users, specifically among one subgroup: alternative client users. Due to Reddit's long-standing extensive and free API, several developers had built alternative mobile apps for Reddit, offering features and designs unavailable on the official Reddit App. Examples of such clients include \textit{Apollo}, \textit{Sync} and \textit{Reddit is Fun}, with Apollo alone reporting close to a million daily active users. The new API pricing strategy rendered these third-party apps economically unfeasible, effectively forcing them out of business \cite{Perez2023May}. Alternative client users who wanted to continue accessing Reddit on a mobile device after the shutdown had to either use Reddit's website or download the official Reddit app, often losing bookmarks, lists and custom settings that had been stored within their third-party app. With many regular Reddit users considering this decision a clear break with the existing concept of Reddit as a relatively open and participatory platform, the announcement caused site-wide protests that, at their peak, were so severe that they caused the site to crash \cite{Peters2023Jun}. Despite these protests, Reddit management held on to its decision, and on July 1st, 2023, all alternative Reddit clients were forcibly cut off from the Reddit backend \cite{eng2024Apr}.  
Reddit had experienced similarly severe protests before in 2015, following the firing of a popular employee and the decision to ban several highly controversial subreddits. The site proved resilient to the protests and lost only a very small number of users \cite{newell2016user}. This time, however, the decision meant shutting off parts of the extended platform infrastructure, potentially directly affecting millions of users. Despite Reddit having 73 million daily active users at the time, considering the known effects of platform retention dynamics, even a comparably small number, such as one or two percent of the user base leaving, could potentially cause problematic cascading effects for the platform \cite{garcia2013social}. 

For this study, we analyze Reddit Data from January 2023 to November 2023, with a special focus on alternative client user groups. Our analysis covers the behavior of alternative app users due to the shutdown on July 1st. It evaluates the willingness to go through with announced threats of leaving and the overall impact of the alternative client depreciation on the platform.

Our results show three main findings. First, while 22\% of alternative client users left the platform after July 1st, only 16\% of alternative client user comment volume was lost. Second, of those users who threatened publicly to leave, about 45\% actually left the platform. Third, while activity among alternative client users significantly decreased, we could find no evidence of an overall effect on Reddit. 
While past research has often found that even comparatively small changes can result in large migrations, our findings indicate that, depending on the circumstances, even highly unpopular and intrusive changes might not result in much activity loss. Whereas before the shutdown, many users and observers had assumed potentially destructive effects for Reddit \cite{Mir2023Jul, zdnet2024Apr}, our data shows that the depreciation of alternative clients did not meaningfully impact the site. While changes to a site's design or structure can have strong negative consequences for platforms, if the best alternative to the original is the same content accessed through a different app, the migration will happen within the platform. Furthermore, we find that the largest contributor to user dissatisfaction was likely not the changes themselves but the incitation by the moderators, who themselves had different motivations for Reddit not to enact the announced changes. This, together with Reddit's perceived offering of unique and interesting content, likely lessened the negative effects of the actual client switch after the fact. 

%% file: sections/2_background.tex
\section{Background and Related Work}

\subsection{Official Reddit App and Third-Party Mobile Clients for Reddit}

The first mobile client for Reddit, \textit{Reddit is Fun}, was launched in 2009 for Android, immediately gaining praise from its users \cite{BibEntry2024Apr}. Subsequently, other clients followed, most notably Sync (2013), Boost (2016), Joey (2017), Apollo (2017) and Infinity (2019). Each client accumulated a significant user base, with the developer of Apollo alone quoting approximately 900,000 daily active users (DAUs) in May 2023 \cite{Perez2023May}. The relevance of these third-party clients for Reddit is highlighted by Reddit's strategic acquisition of an existing app, Alien Blue, in 2014, which the site adopted as its official mobile client. Only in April 2016 Reddit launched a dedicated official Reddit App \cite{Perez2016Apr}. This late entry into the mobile market likely contributed to the wide variety of mobile clients and the habituation of the early user base to these third-party clients.

\subsection{2023 Reddit API Changes}

In April 2023, Reddit announced modifications to its API usage policy, aiming to prevent other companies from using its data to train AI applications, particularly large language models (LLMs) that require extensive textual data \cite{ars023Apr}. From July 1, 2023, the Data API's free usage limits were set to be 100 queries per minute for OAuth-authenticated clients and ten queries per minute for non-OAuth clients, with an option for third-party apps to purchase additional capacity at \$0.24 per 1,000 API calls. On July 5, 2023, access to mature content through the API was to be restricted \cite{reddit2023api}. These changes would have resulted in prohibitive costs for third-party applications like Apollo, making their continued operation economically unfeasible \cite{Perez2023May}. Reddit's decision is seen as part of a broader trend among social media platforms to monetize API access, similar to approaches previously adopted by platforms such as Twitter \cite{Roach2023Jun, Clark2023Jan2}.

\subsection{Migration}

Research on user migration often focuses on the impact of high-friction events such as platform redesigns, de-platforming, blackouts, and policy changes. In the context of deplatforming, much work has focused on how banning hateful communities on Reddit triggered forced migratory movements. \citet{cima2024great} found that after an event titled "The Great Ban" in 2020, around 15\% of affected users left Reddit permanently, a trend also observed in \citet{chandrasekharan2017you}’s study of a similar event in 2015, which found on average approximately 20\% of profiles becoming inactive after banning certain communities. 

Another stream of literature has evaluated the dynamics of migration of users voluntarily leaving a platform due to changes in design, features, or culture. \citet{franzmann2020mobile} evaluated the failed redesign of the popular social media app Snapchat in 2018 and found that the redesign had been so substantial that users had to undergo a new adoption process for the app. Challenging the current assumption that such updates do not affect users' intentions and beliefs if the underlying feature set remains the same, they found that the update negatively influenced users' perceptions of ease of use and caused a strong drop in App Store ratings \cite{fleischmann2016role, amirpur2015keeping}. \citet{gazan2011redesign} analyzed the redesign of the formerly popular Q\&A website Answerbag, finding that many users left due to the removal of popular features hindering social interaction and ensuing interpersonal conflicts about the redesign itself, with alternative websites also specifically targeting potential platform migrants. \citet{wang2024failed} explored the failed migration of academics from Twitter to Mastodon following Twitter's acquisition by Elon Musk, concluding that the absence of a suitable alternative drove users back to their original platform. This acquisition and the following exodus have attracted notable scientific interest. \citet{he2023flocking} found that 2.26\% of Twitter users deleted their accounts as a result of the acquisition, while \citet{jeong2024exploring} observed that users migrating from Twitter to Mastodon often maintained dual-platform usage before ultimately returning to Twitter, challenging earlier predictions of Twitter's collapse. 
The reasons for the success and failure of such voluntary migrations are manifold. \citet{garcia2013social} demonstrated that user departures can disrupt the cost-benefit balance for remaining users, potentially triggering cascading exits with each exit, making the platform less attractive for the remaining users. Besides these push factors, alternative platforms can also incentivize users to jump ship by reducing switching costs and attracting migrating users through targeted incentives \cite{gazan2011redesign, papasavva2024waiting}. 

While the popularity of alternative Reddit clients and the dissatisfaction of their user base about the platform's API monetization decision is well-documented, the actual impact of this decision on the ecosystem, its users, and the platform itself remains unexplored. When looking at the existing literature in the field of user migration, our work offers a bridge between forced migrations due to de-platforming and voluntary migration due to user dissatisfaction. While alternative clients were de-platformed in a way, the actual content and interaction features remained largely the same, albeit in a different look. In this unique position, our study contributes to understanding how users navigate platform-imposed restrictions that do not directly alter the content but significantly change access and user experience. Additionally, we evaluate the behavior of users who openly threatened to leave, providing insight into how such declarations translate into actual migration. As a result, we formulate the following research questions:

\subsection{Research Questions}

\begin{itemize}
    \item \textit{RQ1: Did the shutdown of alternative Reddit apps result in an increased user outflow among third-party client users?}
    \item \textit{RQ2: Were users who publicly announced they would leave the platform more likely to do so compared to those who did not?}
    \item \textit{RQ3: Did the API monetization and subsequent shutdown of alternative clients affect overall Reddit activity?}
\end{itemize}

By evaluating \textit{RQ1}, we aim to better understand whether users actually left the platform or just switched clients. For \textit{RQ2}, we are interested in whether users who publicly announced their disdain were likelier to leave the platform. Lastly, \textit{RQ3} focuses on whether the changes in our sample groups are also reflected in the overall platform activity.

%% file: sections/3_data.tex
\section{Data}

We utilize data from the Arctic Shift dataset by Arthur Heitmann \cite{ArthurHeitmann2024Apr}, sourced through the official Reddit API, excluding private or quarantined subreddits. We take the complete data from r/apolloapp, r/redditsync, r/boostforreddit, r/redditisfun, r/infinity\_for\_reddit, and r/joeyforreddit, from January 1st, 2023 to November 30th, 2023. We then extract the names of all users who made at least one submission in either subreddit during this timeframe (n=9,125). Additionally, we collect all comments posted by these users across Reddit during the same period (n=5,356,382). Following, we filter the data for users who announced they would quit the platform before July 1st, filtering comments for specific n-gram combinations and manually correcting misattributions. Furthermore, we include comment data from the ten most popular subreddits among third-party clients' users, including r/askreddit, r/nba, r/politics, r/nfl, r/worldnews, r/apple, r/soccer, r/technology, r/news and r/baseball.

\subsection{Ethical Considerations}

Our study adheres to strict ethical standards, particularly concerning the sensitive nature of user data. The data is sourced exclusively through the official Reddit API, avoiding private or quarantined subreddits to respect user privacy. In analyzing user behavior, we anonymize identifiable information to prevent personal repercussions for users and minimize false positives through careful filtering of comments for specific n-gram combinations. We also ensure compliance with Reddit API’s terms of service regarding data usage. Our goal is to present our findings responsibly, contributing constructively to discussions on platform governance while maintaining privacy.

%% file: sections/4_methods.tex
\section{Methods}
\subsection{General Approach}
To evaluate the outflow of third-party client users after their discontinuation, we first isolate users of such clients. 
Since Reddit's API does not provide metadata about a user's client, unlike Twitter used to do, we leverage Reddit's unique subreddit structure to identify these users. 
Each third-party client has their own dedicated subreddit for users to discuss the client, suggest improvements and report bugs. 
We assume that people making submissions to these subreddits are also users of the specific client. After conducting a manual review of several hundred comments within the dataset, we are confident in the robustness of the assumption that a user's activity in an alternative client subreddits accurately equates to the same user's utilization of of said third-party client.
We first extract all submissions from the subreddits r/apolloapp, r/redditsync, r/boostforreddit, r/redditisfun, r/infinity\_for\_reddit, and r/joeyforreddit between January 1st, 2023 and November 30th, 2023. In the context of this paper, we define "users" as people interacting with the platform by making comments or submissions. While it would be interesting to also capture the behavior of non-interacting visitors to Reddit, there is currently no reliable and granular data available about this group. 
We extract all users who made at least one submission in either above mentioned subreddit between January 1st, 2023 and November 30th, 2023 and subsequently collect all submissions and comments made by this group during this time frame. Comments and submissions made by AutoModerator where removed from the dataset. While the clients all shut down on July 1st, the subreddits themselves were (and are) still open, with later submissions mostly addressing the loss of features, etc. While these users were not actively using the client at the time of the submissions, we still count them towards the previously active user group. 

\subsection{User Outflow Approach}

We use four distinct methodologies to measure user outflow as a result of the third-party client shutdown. 

\subsubsection{Daily Active Users}
The first approach measures the number of daily active users for each client within our sample. We count towards a client's user base every user who made at least one submission to either client's subreddit between Jan 1, 2023 and Nov 1, 2023.

\subsubsection{User Dropout Rate by Day}
The second approach focuses on measuring the dropout rate by day. For this, we count the number of users per day who made their last comment in the dataset on any given day. As this value naturally increases near its endpoint (culminating to 100\% on the last day captured in the data), we exclude the last month for our analysis to avoid skewed results.

\subsubsection{Absolute Number of Comments Per Day}
For the third approach, we measure the absolute number of comments per day made by each user group. While the first and second approaches only examine user numbers, we use the number of comments to judge whether potentially only casual users left and "power users" with many comments per day remained.

\subsubsection{User Dropout of Users making Threats to Leave}

Additionally, we evaluate the behavior of a subgroup of users that has openly threatened to either abandon or delete their accounts if their favorite third-party client were to shut down. 
We identify these users through a keyword search of our initial user sample, followed by manual verification to exclude false positives. We classify every user as "making a threat" that strongly implied or directly announced that they would abandon Reddit if the platform enacted the API changes. 
Examples of positive matches include \textit{"Without Apollo I’m likely to leave Reddit"} and \textit{"if [Apollo] gets removed then goodbye reddit"}. Very vague threats such as \textit{"I don’t know what I’ll do without apollo. I want to say I’ll quit reddit but they got me by the balls I think."} were not counted.

\subsection{User Engagement}

To understand the impact on the overall platform, we also evaluate activity metrics for all Reddit users in a sample of subreddits. Thus, we can contrast our sample with all of Reddit and better understand potential platform-wide changes.

\subsubsection{Subreddit Activity}

To assess the maximum potential effects the API changes could have had, we select the 10 most popular subreddits among our sample of third-party client users, excluding the subreddits for the third-party clients themselves \textit{(r/askreddit, r/nba, r/politics, r/nfl, r/worldnews, r/apple, r/soccer, r/technology, r/news and r/baseball)}. We track the daily number of comments among all Reddit users. If the third-party client shutdown had any global effects, they should be reflected in the selected sample of subreddits.

\subsubsection{New User Account Creation}

Considering the possibility of users deleting their accounts in protest and later returning to the platform with a new account, we examine the influx of new users. 
As account creation dates are unavailable, we instead analyze the number of first-time commentators for the ten most popular subreddits among third-party client users between January and November 2023. A noticeable increase of people making their first comment after July 1st would indicate a return of users that previously left the platform.

%% file: sections/5_results.tex
\section{Results}

\subsection{Daily Active Users}

\begin{figure}[H]
   \centering
   \includegraphics[width=\columnwidth]{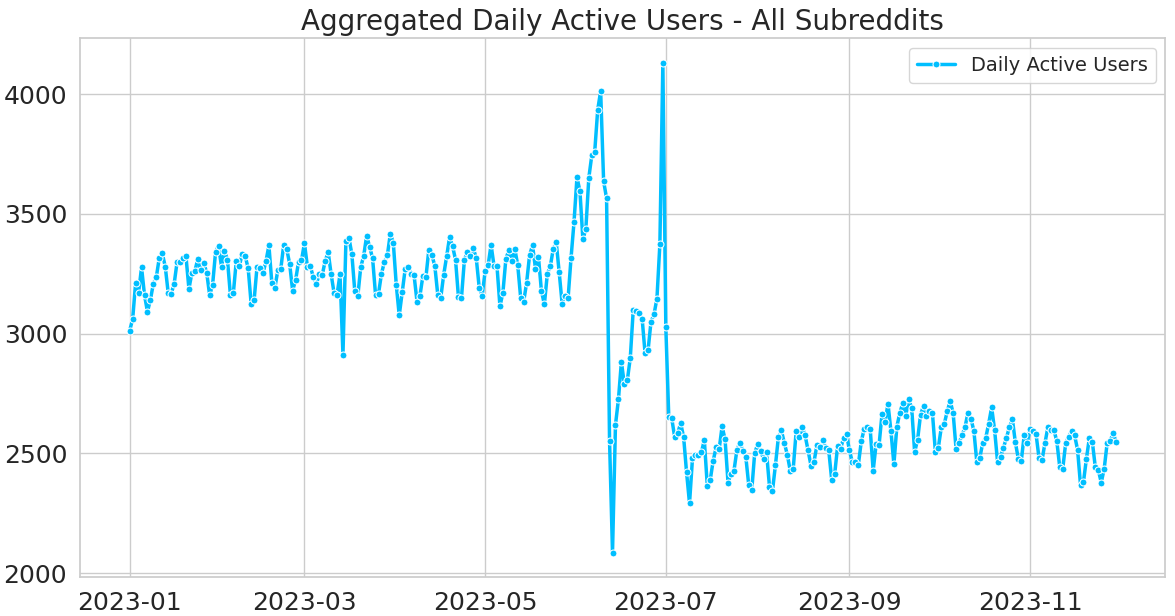} 
   \caption{Daily Active User Numbers for All Clients}
   \label{fig:dau_total}
       \vspace{-0.5em} 
\end{figure}

\begin{figure}[H]
   \centering
      \vspace{-0.5em} 
   \includegraphics[width=\columnwidth]{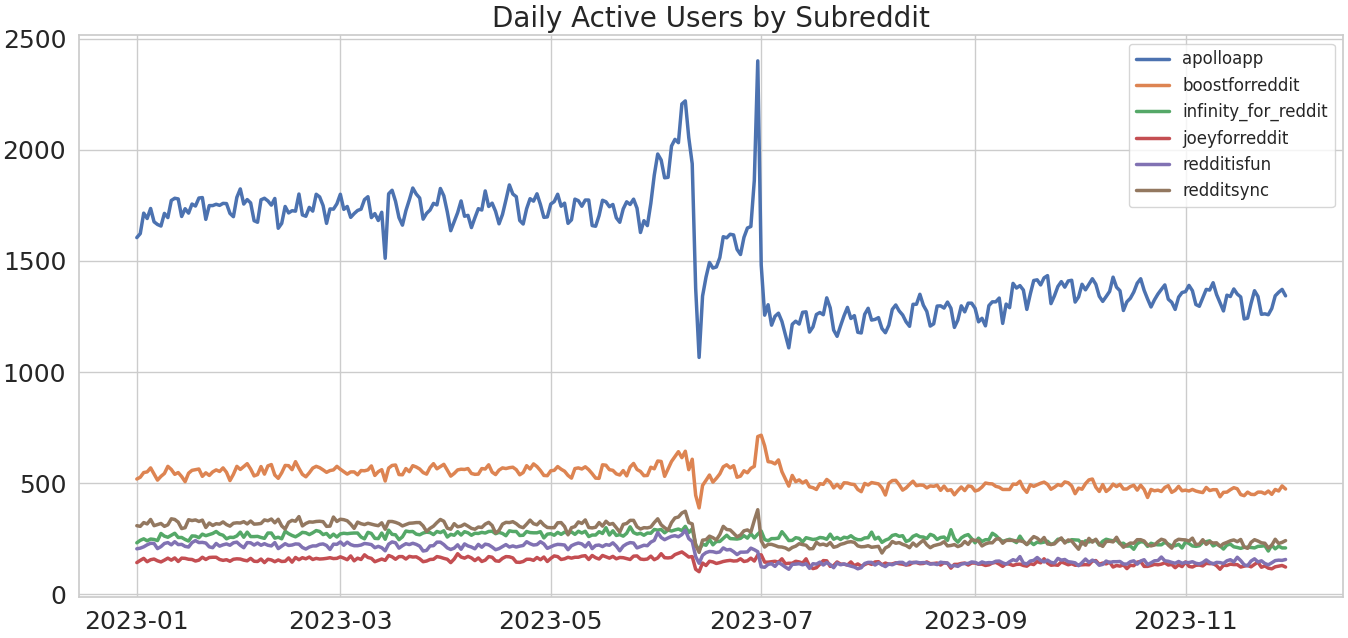} 
   \caption{Daily Active User Numbers by Used Client}
   \label{fig:dau_by_subreddit}
\end{figure}

Displayed in Figure \ref{fig:dau_total} and Figure \ref{fig:dau_by_subreddit}, the DAU numbers experience a significant drop-off in the week following June 30th across all observed client user groups. Prior to this, during the mid-June protests, a considerable number of users either deleted or abandoned their accounts. In most cases, the numbers briefly returned to normal or close-to-normal levels in sync with the general protest course before June 30th. A comparison between the average number of DAUs in the four months preceding (Feb/Mar/Apr/May) and the four months following (Aug/Sept/Oct/Nov) the protests and API monetization in June and July revealed a significant decrease in users across all clients, as shown in Table \ref{tab:DAU}. The number of DAUs decreased by 22\% after the alternative client shutdown across the users of these clients.

\begin{table*}[t]
\begin{center}
\begin{tabular}{|l|l|l|l|}
\hline
Subreddit & Before Strike & After Strike & Pre-Post-Comparison \\
 & (Feb/Mar/Apr/May) & (Aug/Sept/Oct/Nov) & \\
\hline
Apollo
    & \textit{M}(\textit{SD}) = 1,736.35 (52.68)     
    & \textit{M}(\textit{SD}) = 1,322.26 (60.25)     
    & \textit{t}(240) = 56.88, \textit{p} $<$ .001   $\searrow$ \\
Boost for Reddit
    & \textit{M}(\textit{SD}) = 558.05 (17.63)     
    & \textit{M}(\textit{SD}) = 478.51 (17.32)     
    & \textit{t}(240) = 35.41, \textit{p} $<$ .001   $\searrow$  \\
Infinity\_for\_Reddit
    & \textit{M}(\textit{SD}) = 272.3 (10.71)     
    & \textit{M}(\textit{SD}) = 236.54 (18.52)     
    & \textit{t}(240) = 18.42, \textit{p} $<$ .001   $\searrow$  \\
Joey for Reddit
    & \textit{M}(\textit{SD}) = 160.97 (8.46)     
    & \textit{M}(\textit{SD}) = 134.51 (7.94)     
    & \textit{t}(240) = 25.09, \textit{p} $<$ .001   $\searrow$  \\
Reddit is Fun
    & \textit{M}(\textit{SD}) = 220.14 (9.99)     
    & \textit{M}(\textit{SD}) = 144.51 (8.7)     
    & \textit{t}(240) = 62.85, \textit{p} $<$ .001   $\searrow$  \\
Sync
    & \textit{M}(\textit{SD}) = 315.05 (14.74)     
    & \textit{M}(\textit{SD}) = 231.66 (13.31)     
    & \textit{t}(240) = 46.21, \textit{p} $<$ .001   $\searrow$  \\
\hline
Total
    & \textit{M}(\textit{SD}) = 3262.86 (87.94)     
    & \textit{M}(\textit{SD}) = 2547.99 (84.53)     
    & \textit{t}(240) = 64.48, \textit{p} $<$ .001   $\searrow$  \\
\hline
\end{tabular}
\end{center}
\caption{Between Group Comparisons for the differences in mean number of daily active users in the four months before (Feb/Mar/Apr/May) and the four months after (Aug/Sept/Oct/Nov) the strike- and API-monetization-month per client user group and in total. Student's T-Tests were conducted for all tests except 'Infinity\_for\_reddit', for which a Welch Test was used due to a violation of the variance homogeneity assumption. As one test per subreddit was performed, all p-values were Bonferroni-adjusted to p=.05/6 for the subreddit tests.}
\label{tab:DAU}
\end{table*}

\subsection{User Dropout by Day}
\begin{figure}[h]
   \centering
   \includegraphics[width=1\columnwidth]{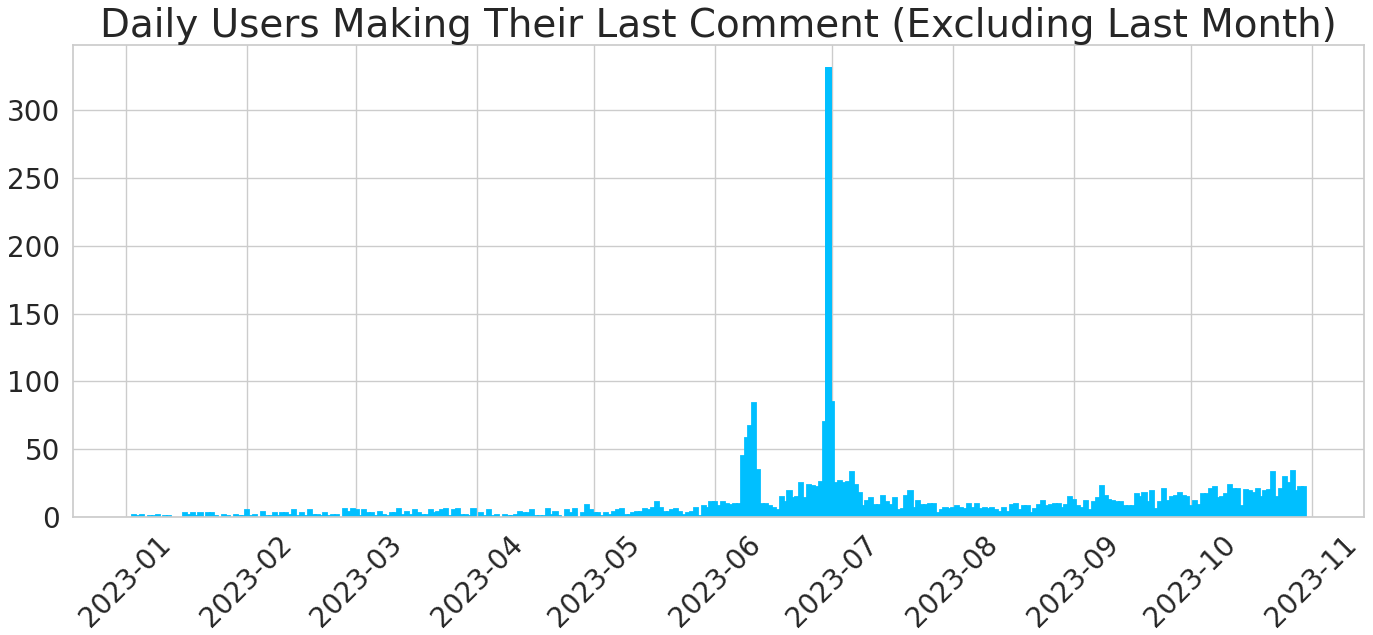}
   \caption{User Dropout by Day}
   \label{fig:user_dropout_by_day}
\end{figure}

Further explorations of user dropout by day reveal a large spike on June 30th, 2023, with 331 users making their last comment on this day alone, which sharply contrasts with the daily average of 6.17 users over the previous six months (see Figure \ref{fig:user_dropout_by_day}). Another 4\% of users in the data set deleted or abandoned their Reddit account in the 14 days after the client shutdown between July 1st and July 14th, 2023. 

\subsection{Daily Comments}

\begin{figure}[H]
   \centering
   \includegraphics[width=1\columnwidth]{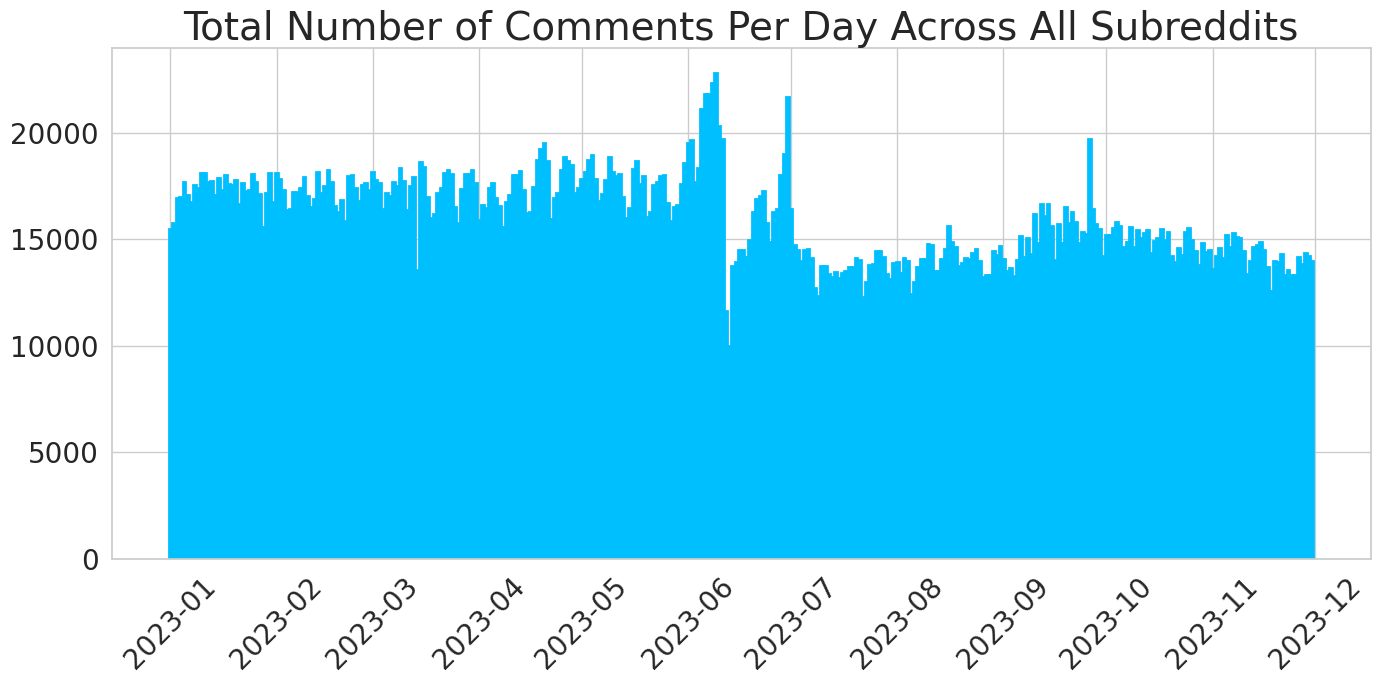}
   \caption{Daily Comments by Third-Party Client Users}
   \label{fig:daily_comment_third_party_client}
\end{figure}

Comments made by third-party client users follow a similar pattern (Figure \ref{fig:daily_comment_third_party_client}). 
There is a large spike before the June 12th protest and a sharp drop-off during the protests themselves. 
Similarly, a surge of comments in the week preceding the API monetization is visible. 
Overall, after the strike and API changes, third-party client users made on average 16\% fewer comments per day compared to the period before the changes ($M_{Feb-May}$ = 17,403.57, $M_{Aug-Nov}$ = 14,609.74, $SD_{Feb-May}$ = 917.03, $SD_{Aug-Nov}$ = 974.72, \textit{t}(240) =22.96, \textit{p}  $<$ .001).

\subsection{Users Threatening to Leave}

\begin{figure}[h]
    \centering
    \includegraphics[width=1\columnwidth]{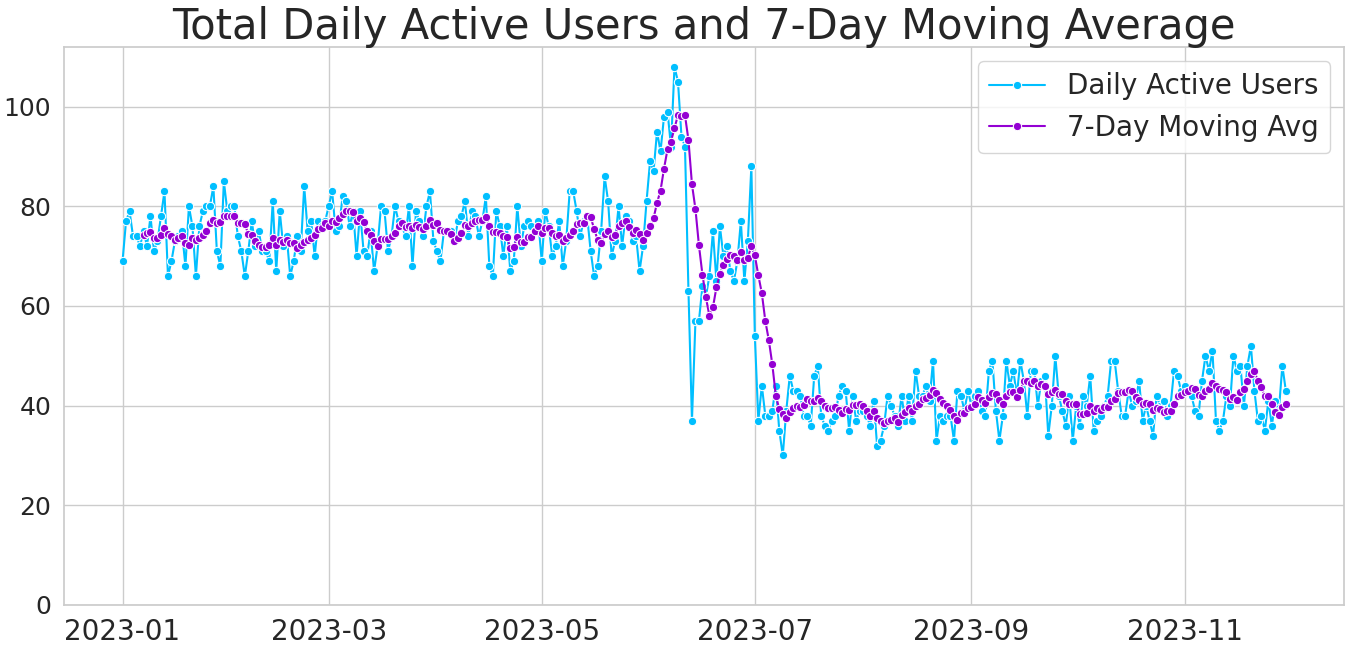}
    \caption{Number of Daily Active Users That Threatened to Leave Reddit}
    \label{fig:dau_threaten}
\end{figure}

Approximately half of the users who openly threatened to leave the platform actually ceased using their accounts or deleted them on July 1st, with a smaller subgroup discontinuing their activities during the earlier protests (Figure \ref{fig:dau_threaten}).
Comparing the mean number of daily active users in the four months before and after the strike and API monetization reveals that the number of daily active users decreased by 45\% ($M_{Feb-May}$ = 74.92, $M_{Aug-Nov}$ =  41.18, $SD_{Feb-May}$ =  4.62, $SD_{Aug-Nov}$ =  4.6, \textit{t}(240) = 56.89, \textit{p}  $<$ .001). Furthermore, the mean number of daily comments by these users decreased by 41\% ($M_{Feb-May}$ = 463.18, $M_{Aug-Nov}$ =  272.57, $SD_{Feb-May}$ =  52.44, $SD_{Aug-Nov}$ =  50.64, \textit{t}(240) = 28.76, \textit{p}  $<$ .001).

\subsection{Effects on Global Platform Activity}
\begin{figure}[h]
    \centering
    \includegraphics[width=1\columnwidth]{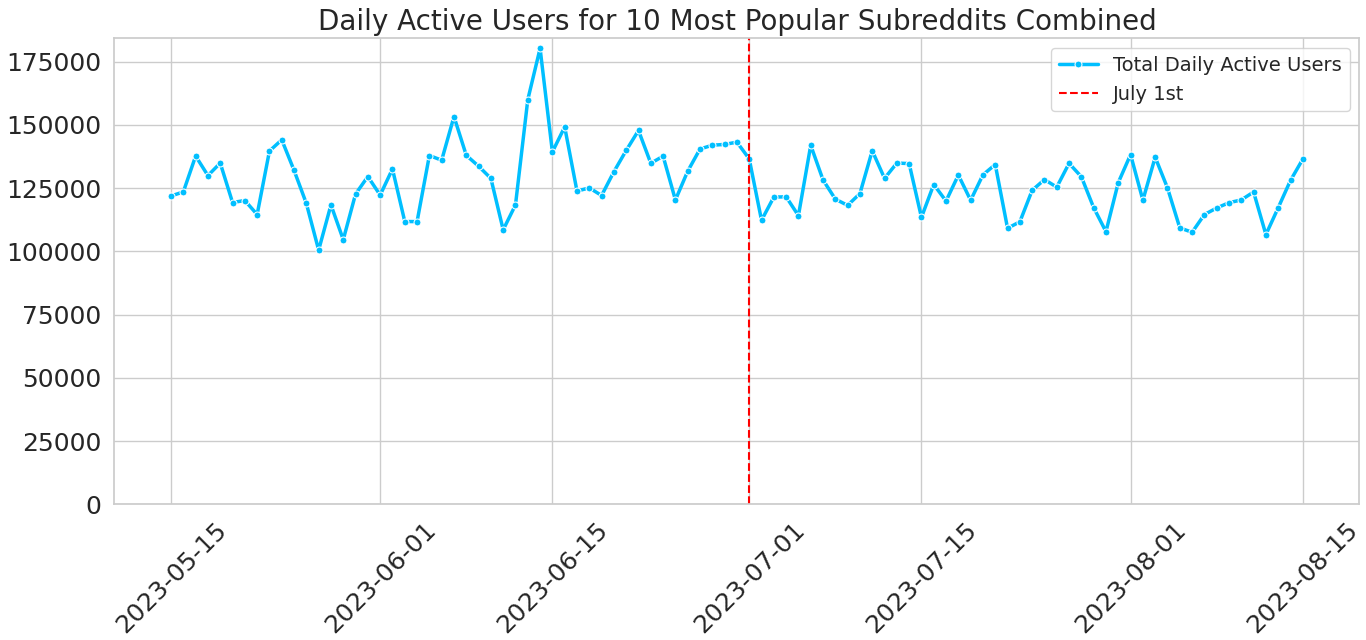}
    \caption{Effects on Global Platform Activity}
    \label{fig:aggregated_total}
\end{figure}

We also analyze potential effects on subreddit usage to examine whether the shutdown affected platform usage. This analysis aims to identify potential changes that our initial sample selection might not have covered. For our analysis, we count the number of Daily Active Users (DAUs) in the 10 most popular subreddits among our sample of third-party client users. If the shutdown substantially affected the platform, it would be most evident within these subreddits.
However, our findings, as visualized in Figure \ref{fig:aggregated_total}, indicate stability in subreddit usage, with no significant decline in the number of DAUs beyond normal fluctuations in the 4 months before and after the strike and API shutdown ($M_{Feb-May}$ = 127,199.93, $M_{Aug-Nov}$ =  126,724.84, $SD_{Feb-May}$ =  10,823.57, $SD_{Aug-Nov}$ =  11,301.67, \textit{t}(240) = .33, \textit{p}  = .738). 

\subsection{New Account Creation}

\begin{figure}[h]
    \centering
    \includegraphics[width=1\columnwidth]{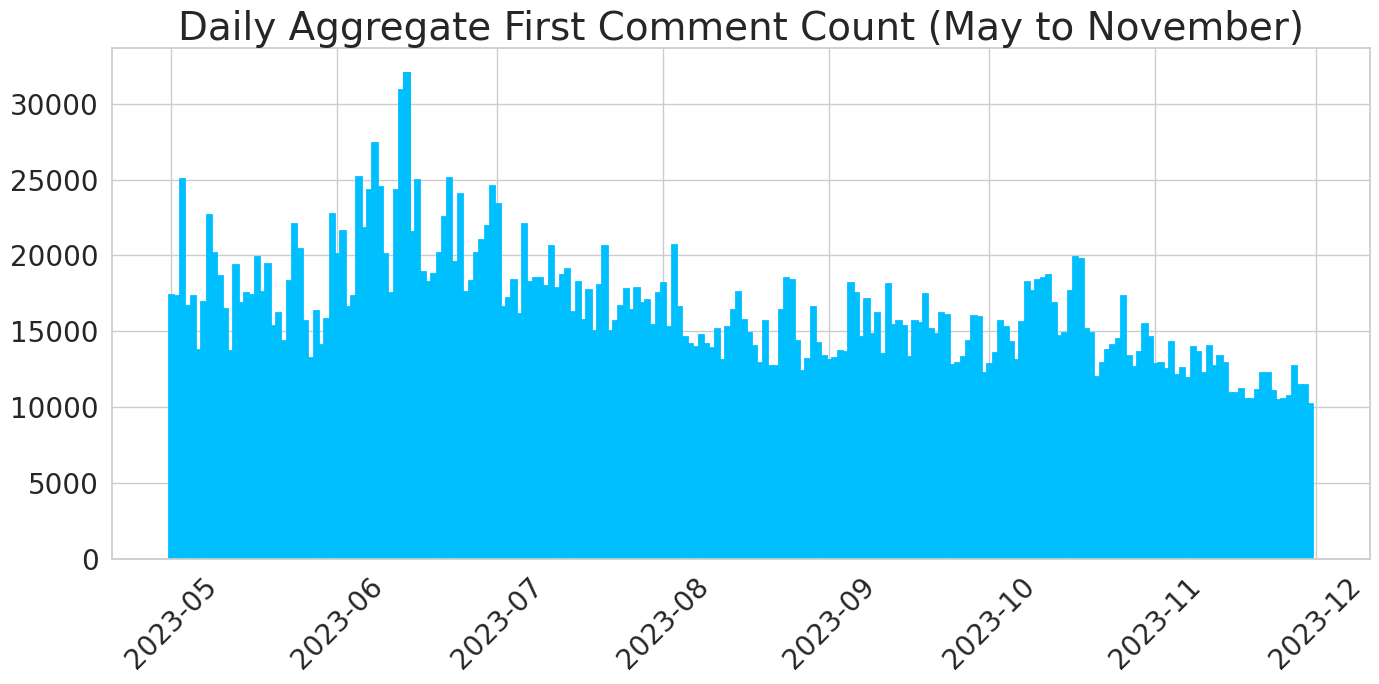}
    \caption{Number of first-time commentators between May and November 2023 for the 10 most popular subreddits among alternative client users}
    \label{fig:first_comment}
\end{figure}

Examining the number of accounts that made their first comment on any given day in the ten subreddits most popular with third-party client users indicates two main developments. While the protests in mid-June caused many people to access various new subreddits and make comments in them, in the weeks following the shutdown of alternative apps, our analysis shows no significant changes in the behavior of new commentators.
Due to the nature of our data collection methodology, an initial incline in first-time commentators is expected. On the first day of data collection, the number of users making their first comment is naturally higher. Over time, however, this number stabilizes and becomes more indicative of new user activity, particularly after the July 1st cutoff.  
Hence, for a better representation, Figure \ref{fig:first_comment} only shows data beginning in May.

%% file: sections/6_discussion.tex
\section{Discussion}

The depreciation of alternative Reddit clients following July 1, 2023 had no discernible effects on the platform. While the subgroup of alternative client users saw significant decreases in user and comment activity, this did not cause cascades of other users to leave. This outcome contradicts the concerns previously raised by users, developers, and observers before the shutdown, who warned that such measures could deter existing users and potentially harm Reddit financially and structurally \cite{Mir2023Jul, zdnet2024Apr}.

Our results offer new insights into the effects of migration at the intersection between forced migrations due to deplatforming and voluntary migration due to user dissatisfaction. In literature, many observed migratory events in online spaces motivated by user dissatisfaction have had strong negative consequences for the affected platform. However, in the context of forced migration due to deplatforming (such as banning only particularly hateful subreddits), only small numbers of users usually completely abandon the platform, with most just continuing their activity in different subreddits. The question arises as to why, in this case, the strong dissatisfaction of users about the pending changes did not result in strong negative effects within the affected user group. 

One potential explanation for our results in this context lies within the unique structure of Reddit. Reddit is organized in subreddits, individually themed forums with their own moderators, rules and community etiquette. Due to the way subreddit moderators are appointed, many moderators have held their positions for a considerable time and are invested in the platform. As mentioned in the beginning, most regular users likely would not have cared much about changes in their user interface. When Twitter cut off third-party clients, this was discussed in the media but caused no further notable disruptions or actions by users on the platform itself \cite{Singh2023Feb}. 

The protests on Reddit were mainly motivated by subreddit moderators, who used their powers as gatekeepers to block users from accessing their subreddits. Notably, the developer of Apollo himself took a leading role in the protests \cite{Davis2023Jun}. It seems unlikely that without the moderator's efforts, similar strong effects on platform activity during the protests would have been possible. The main drivers for the protests and their most severe effects, therefore, were effectively a relatively small group of individuals that had the highest interest in continuing to influence the platform's direction. While the vast majority of users would not have been directly affected by the changes in API pricing and subsequent client shutdowns, they were, however, affected by the protests initiated by moderators, barring access to a large number of communities and forcing users to switch to alternative subreddits, causing uncertainty among the user base.

This offers a plausible explanation for the comparably small effects of the changes on the alternative client user base as well as Reddit after July 1st compared to the massive disruptions caused by the protests beforehand. With the protests not mainly driven by affected users being angry about the actual shutdown of the alternative clients but more so by site moderators feeling unappreciated, a seemingly large protest might not have had many users truly supporting the cause behind it. While some alternative client users actually left, the vast majority has remained active on Reddit's official offerings. This also falls in line with earlier findings, which support the notion that the user perception of design changes should not influence their intention to use the service as well as actual usage behavior as long as the underlying features remain the same \cite{amirpur2015keeping, fleischmann2016role}.

Previous research has emphasized the importance of Reddit both as vast knowledge base and home for many unique and niche communities, characteristics that may contribute to a particularly high stickiness of the platform and the subsequently relatively small effects observed in our findings \cite{newell2016user}. There are certain clues however that lead us to believe that the underlying mechanisms influencing user migration are still generally comparable to those seen on platforms that lack Reddit’s vast collection of niche content \cite{garcia2013social}. For example, during the mass protests preceding the alternative client shutdown, several communities successfully moved altogether to different platforms \cite{rehab2024}. Furthermore, the overwhelming majority of Reddit users only very rarely comments on submissions older than a couple of days. While Reddit's existing collection of knowledge likely does lead many people to the site via search engines, these visitors do not necessarily positively influence the creation of new content or steadiness of existing communities. Just like for any other social media site, if enough users move to a new place, the cost/benefit balance can become skewed for existing users \cite{he2023flocking}. While Reddit undoubtedly possesses a certain stickiness, there is no evidence pointing towards Reddit being particularly irreplaceable compared to other social media platforms such as Snapchat, Facebook or Twitter. 

For Reddit itself, the changes likely had favorable financial outcomes. Alternative Reddit apps show no Reddit ads but instead display their own ads that benefit the app developers. Therefore, losing individual users came at no direct cost for Reddit, while every user motivated to switch generated additional income. Even if all alternative app users had quit as a result of the shutdown, no advertisement income would have been lost. In 2023, Reddit had \$804mln in revenue and 73.1 million daily active users (DAUs). Reddit, on average, earned \$12,31 per daily active user per year with advertisement \cite{sec2024Feb}. Looking at our observed retention rates, with 78\% of users remaining active on the platform, shutting down Apollo with its claimed 900.000 DAUs alone earned Reddit an additional \$8-\$9mln in advertisement revenues annually. With all other observed clients together showing approximately the same DAU numbers as Apollo, we estimate the overall gains for Reddit to be between \$16-18mln. Furthermore, Reddit has successfully sold API access to companies training LLMs, with a deal volume of already over \$200mln over the next three years \cite{arstech2024Feb}. 

Our data ends in November 2023, with long-term effects possibly not yet showing. During the 2015 protests on Reddit, considerable numbers of users became "tourists" or "dual citizens" on alternative platforms \cite{newell2016user}. This behavior could also be observed this time, with some moderators migrating their communities during the protests to alternatives such as Lemmy, Discord or Raddle \cite{rehab2024}. Lemmy specifically saw its active user numbers rise drastically from 1,385 to about 68,947 between June and July 2023; these numbers, however, decreased in the following months to around 32,000 and only later started rising again \cite{Fedi2024Apr}. Data from Google Trends supports this notion, showing that interest in the terms "Mastodon" and "Fediverse" more than tripled in the week following July 1 before quickly returning to previous levels in the subsequent weeks \cite{googletrends2023}. These numbers are considerable for small platforms such as Lemmy, but negligible for Reddit. While the shutdown of third-party clients was certainly a nuisance for many users, it did not trigger a negative cost-benefit cascade with increasing numbers of users leaving. As the main migratory goal for most people was still Reddit, just within another app, the vast majority of users and content creators remained active on the platform. 

\subsection{Limitations and Future Work}

The results in this paper come from a single incident on Reddit. While the scale of the incident was quite large, this potentially limits the applicability of the results to other platforms. Additionally, the sample used in this paper does not necessarily represent the full subset of third-party client users due to the nature of its collection. A considerable chunk of Reddit's traffic is made up by "lurkers" which are visiting the platform but not leaving comments or creating submissions \cite{Nielsen2022Apr}. Our data collection method is unable to capture the behavior of this group. As alternative clients however were primarily designed with features catering to power users, it is likely that only a small group of "lurkers" utilized them. The significant changes in user behavior in the days after the shutdown and the congruence of the results when measuring the overall effects across the whole population are indicators of a good sample selection. Another limitation of our study is the inability to track individual users across multiple platforms, which reduces the explanatory power of our findings compared to analyses based on per-user tracking. Additionally, we are unable to distinguish between comments made through third-party clients and those made via web browsers on desktop computers. Furthermore, a small group of users later discovered ways to continue using some of the observed alternative clients beyond July 1 by opening an empty subreddit to obtain moderator API access keys. Due to the complicated nature of both obtaining the keys and patching the affected apps, the group of users still using alternative apps, however, can be considered insignificant for this study. An interesting avenue for future work could also be an extended observation of migratory events on platform dynamics and user engagement, exceeding five-month period in this paper. 

\subsection{Conclusion}

Overall, our analysis indicates that Reddit's decision to monetize its API, leading to the shutdown of alternative clients, had a very limited impact on the platform overall. While small numbers of users left, the engagement loss among alternative client users was minor and left no measurable impact on the platform overall. Judging from our findings, the loss of alternative clients was, after all, not a migration-triggering negative event for most users. However, the earlier incitation of users by moderators during the protest had likely amplified the perceived size of the issue within the user base and caused uncertainty beforehand. Reddit's previously highlighted unique and interesting content structure, coupled with the fact that switching costs for users towards a new app were in all cases lower than switching to an entirely different site such as Mastodon, resulted in many users just accepting a client switch, helping to retain the platform's user base, even in the face of unpopular changes.